\documentclass[5p, twocolumn]{elsarticle}

\usepackage[english]{babel}
\usepackage[T1]{fontenc}
\usepackage[utf8]{inputenc}
\usepackage{amssymb}
\usepackage[hidelinks]{hyperref}
\usepackage{longtable}
\usepackage{graphicx}
\usepackage{booktabs}
\usepackage{amsmath}

\makeatletter
  \renewcommand{\fps@figure}{htb}
  \renewcommand{\fps@table}{htb}
\makeatother 
   
\begin{document}
\title{Investigation of the effects of valve closing in a static expansion system}

\author[1]{Th. Bock\corref{cor1}}
\ead{thomas.bock@ptb.de}

\author[1]{M. Bernien}
\ead{matthias.bernien@ptb.de}

\author[1]{Ch. Buchmann}
\ead{christian.buchmann@ptb.de}

\author[1]{T. Rubin}
\ead{tom.rubin@ptb.de}

\author[1]{K. Jousten}
\ead{karl.jousten@ptb.de}

\cortext[cor1]{Corresponding author}

\address[1]{Physikalisch--Technische Bundesanstalt, Abbestraße 2--12, 10587 Berlin, Germany}

\begin{abstract}
 The precise measurement of the pressure of the gas enclosed in the
 starting volume of a static expansion system is important for
 achieving low measurement uncertainties. This work focuses on the
 valve closing process and the resulting pressure difference. The condition of flow
 conservation is used to set up a model and an experimental
 procedure has been developed to investigate the pressure change
 caused by the moving valve plate. Various parameters are evaluated
 and compared with experimental data.
\end{abstract}


\maketitle
\section{Introduction}
\label{sec:intro}

To this day static expansion systems are one of the most accurate
realizations of the pressure scale in the range from
$1\times10^{-2}$\,Pa to $100$\,Pa and are used as primary standards in
many metrology institutes~\cite{ricker}. Many contributions to the
uncertainty have been investigated in great detail, e.\,g., the impact
of the temperature during expansion and real gas
effects~\cite{jousten94, jitschin}.  In this study we focus on the
dynamic effects during the closing process of the valve that encloses
the gas to be expanded. These effects become increasingly important
when working with small volumes.

The PTB vacuum group currently evaluates the new static expansion
primary standard SE3. SE3 will be a fully automated calibration
facility for the realization and dissemination of pressures in the
range $1\times10^{-2}$\,Pa to $100$\,Pa. Six DN16CF valves described
in this paper are used for three starting volumes with nominal volumes
of $0.02$\,l, $0.2$\,l and $2$\,l. The calibration vessel has a volume
of $200$\,l. The entire calibration pressure range will be covered by
three different single stage expansions. This is beneficial since
multiple consecutive expansions lead to increased uncertainties.  The
principle of static expansion \cite{wutz} relies on the transfer of a
fixed amount of gas from a small volume into a larger volume. If the
initial pressure and the ratio of the two volumes is known precisely,
a well-known lower pressure is generated.  In order to generate a
calibration pressure of $1\times10^{-2}$\,Pa an expansion ratio
$f=V_\mathrm{s}/(V_\mathrm{s}+V_\mathrm{e})$ of $1\times10^{-4}$ and a
filling pressure $p_\mathrm{fill}$ of $100$\,Pa is used. $V_\mathrm{s}
= 0.02$\,l is the smallest starting volume of SE3. $V_\mathrm{e}$ is
the $200$\,l volume of the calibration vessel. The generated pressure
$p_\mathrm{after}$ is obtained using the ideal gas law and the
condition that the amount of gas is constant:
\begin{align}
	\frac{V_\mathrm{s}p_\mathrm{before}}{T_\mathrm{before}}&=N k_\mathrm{B}
	=\frac{(V_\mathrm{s}+V_\mathrm{e})p_\mathrm{after}}{T_\mathrm{after}}\\
	\Rightarrow\quad
	\label{equ:pgen}
	p_\mathrm{after}&=p_\mathrm{before}\frac{V_\mathrm{s}}{V_\mathrm{s}+V_\mathrm{e}}
	\nonumber \frac{T_\mathrm{after}}{T_\mathrm{before}}
	=(p_\mathrm{fill}+\Delta p)f\frac{T_\mathrm{after}}{T_\mathrm{before}}\\
	&=p_\mathrm{fill}\Big(1+\frac{\Delta p}{p_\mathrm{fill}}\Big)
	f\frac{T_\mathrm{after}}{T_\mathrm{before}}
\end{align}
with $k_\mathrm{B}$ the Boltzmann constant, $N$ the number of gas molecules,
$p_\mathrm{before}$ the pressure and $T_\mathrm{before}$ the temperature of
the enclosed gas before expansion as well as $T_\mathrm{after}$ the temperature of
the gas after expansion. The filling pressure $p_\mathrm{fill}$
can only be measured outside the starting volume
since a gauge introduces heat and additional volume. $\Delta p$
denotes a potential pressure difference between the position of the
gauge and inside of the closed starting volume. This pressure
difference is the main topic of this work.

After presenting an initial investigation, the layout and the design
of the starting volumes as well as the associated valves are described in
the following. Then the pressure change during the closing phase
is modeled. The derived dependencies are compared to experimental
results. The experiments were performed with two different valves of
the same type. Expansion ratio measurements with nitrogen, helium and
argon are presented. A correction factor and an additional measurement
uncertainty are introduced.

\section{Initial investigation}
\label{sec:ini}
\begin{figure}
    \centering
    \includegraphics[angle=0, width=\columnwidth]{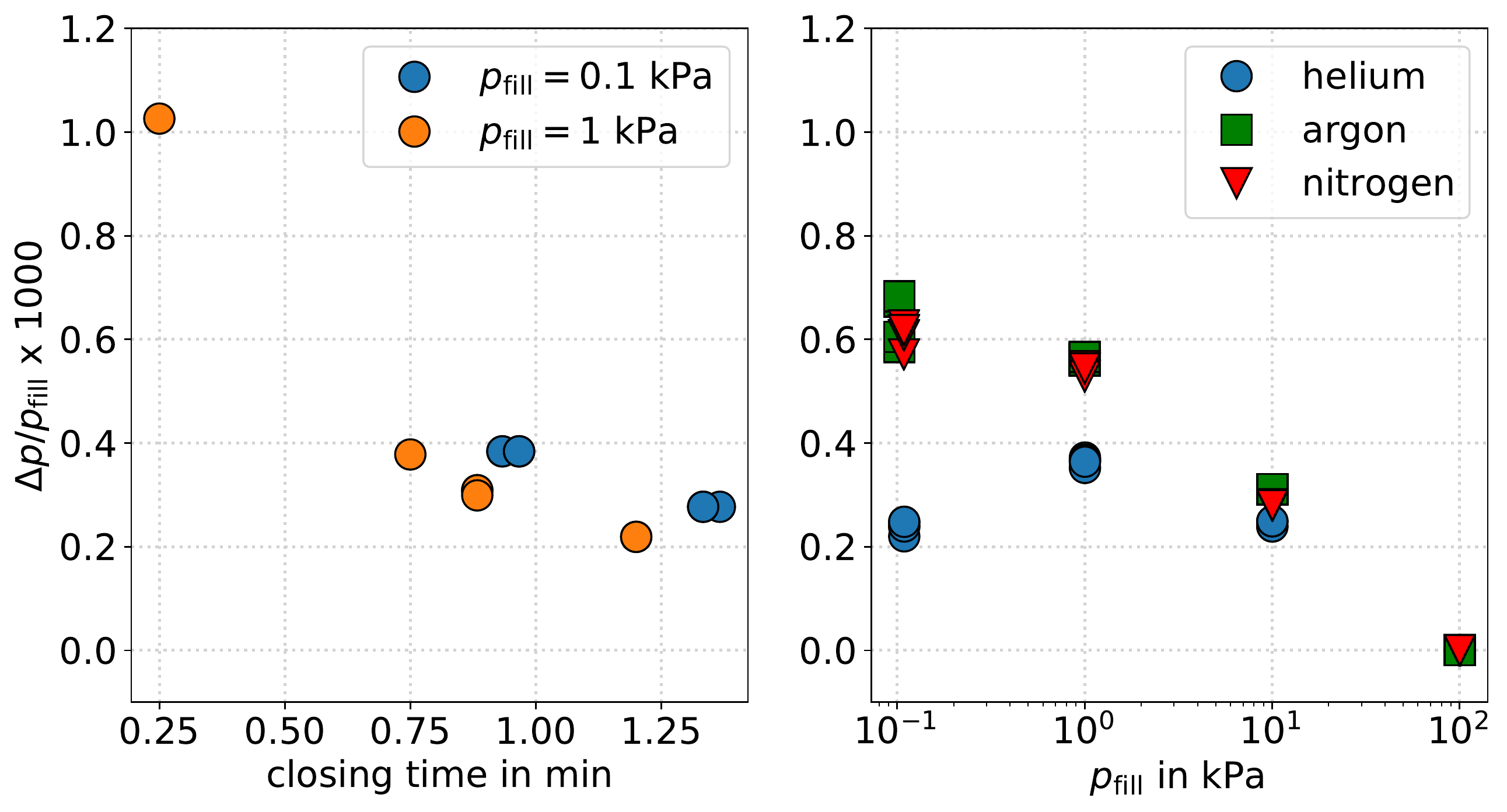}
    \caption{Results of initial investigations with a pneumatic DN16CF
      angle valve of the type \emph{VAT~57124--GE44--0002} and an
      experimental setup as shown in Figure~\ref{fig:exp_scheme}. The
      relative pressure differences across the valve plate after the
      closing process are plotted as a function of closing time for
      two different filling pressures of helium (left) and as a
      function of filling pressure for helium, argon and nitrogen with
      a closing time of 1\,min (right). The volume enclosed by the
      valve was about 30\,cm$^3$.  The closing time was varied via a
      throttle and relates here to the entire valve stroke of about
      $7$\,mm. 1\,min closing time corresponds to a closing speed of
      about 120\,$\mu$m/s.}
    \label{fig:pre}
\end{figure}

To check for pressure differences due to valve closing of the starting
volume, initial investigations with two pneumatic DN16CF all-metal
angle valves of the type \emph{VAT~57124--GE44--0002} and
\emph{VAT~F57--74423--02} were conducted. The latter is modified: it
has a reduced closing path of $2$\,mm and is used on the primary
standard SE1 \cite{jousten93, jousten99}. Figure~\ref{fig:pre}
shows results with the valves of the first type. A
differential capacitance diaphragm gauge (CDG) was used to measure
$\Delta p$ after closing the valve while varying the filling pressure,
closing speed and gas species (see Figure~\ref{fig:exp_scheme}).
The experiments show that after the valve's closing
position was reached a constant pressure indication remains.  Relative
differences up to $1\times10^{-3}$ occur at high closing
speeds. Varying the filling pressure causes relative differences up to
$7\times10^{-4}$, depending on the gas species.  It became obvious that
these deviations would significantly increase the measurement
uncertainty of the system. To gain a better understanding, time
resolved experiments were carried out as described
below. The use of servomotors allows for a precise speed and force
control of the valve closing process.

\section{Valves of the starting volume}
\label{ssec:valves}

\begin{figure}
  \centering
  \includegraphics[angle=0, width=\columnwidth]{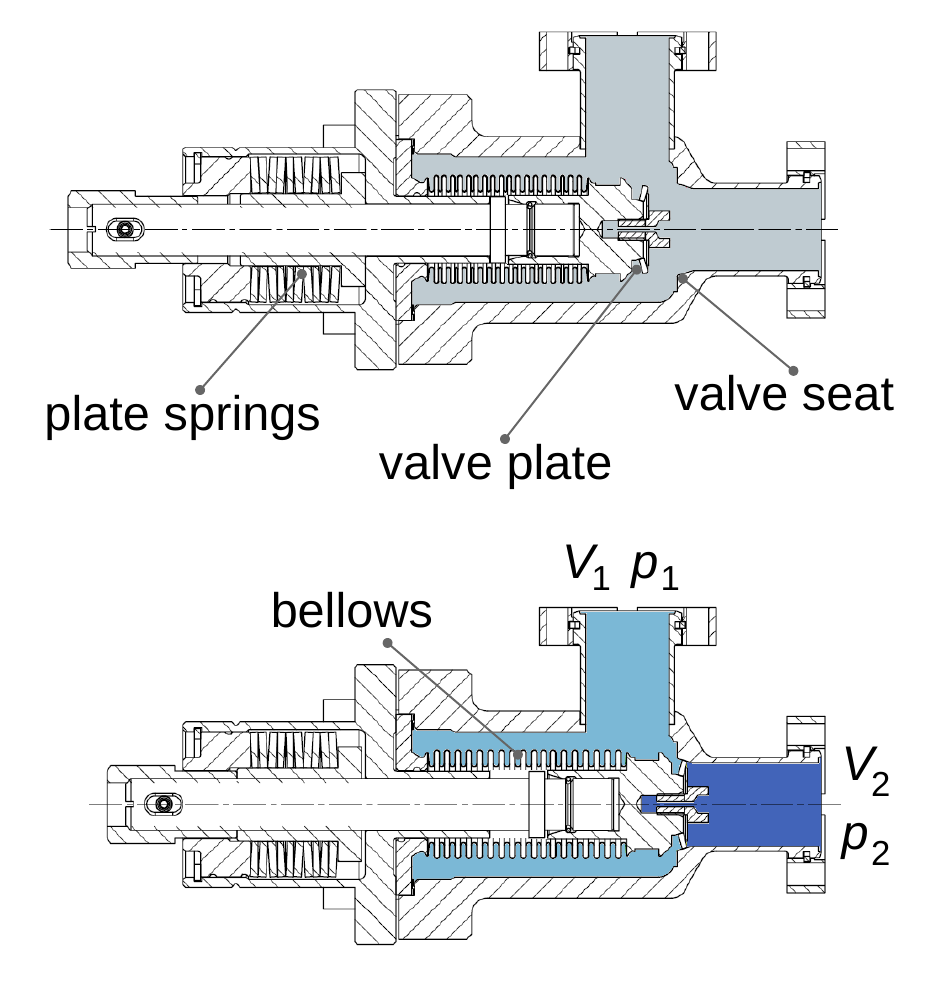}
    \caption{Schematic drawing of the valve closing
      mechanism.}
    \label{fig:valve_scheme}
\end{figure}

Each of the starting volumes of SE3 consists of an aluminum cylinder
and two DN16CF valves (inlet and outlet valves). The aluminum
cylinders are designed and build to ensure an easily measurable
volume. With a nominal volume of $5$\,cm$^3$ the valves contribute
with $50$\,\% to the value of $V_\mathrm{s}$.  The inlet and the outlet valves
used for $V_\mathrm{s}$ are DN16CF all-metal angle valves of the manufacturer
VAT (Type 571 24-GE02).
A schematic drawing of the valves is shown in Figure~\ref{fig:valve_scheme}.
The valve plate is a sealing ring mounted to a piston with a bellow as a
flexible element that is mechanically coupled to a threaded rod. Upon rotation
the threaded rod moves forward in a hollow shaft. When reaching the closing position
the sealing ring seals against the valve seat and the piston. The force acting on
the sealing ring is lifting the hollow shaft and transferred to plate springs that
provide a well-defined closing force.

\subsection{Geometrical dimensions}

\begin{figure*}
    \centering
    \includegraphics[angle=0, width=\linewidth]{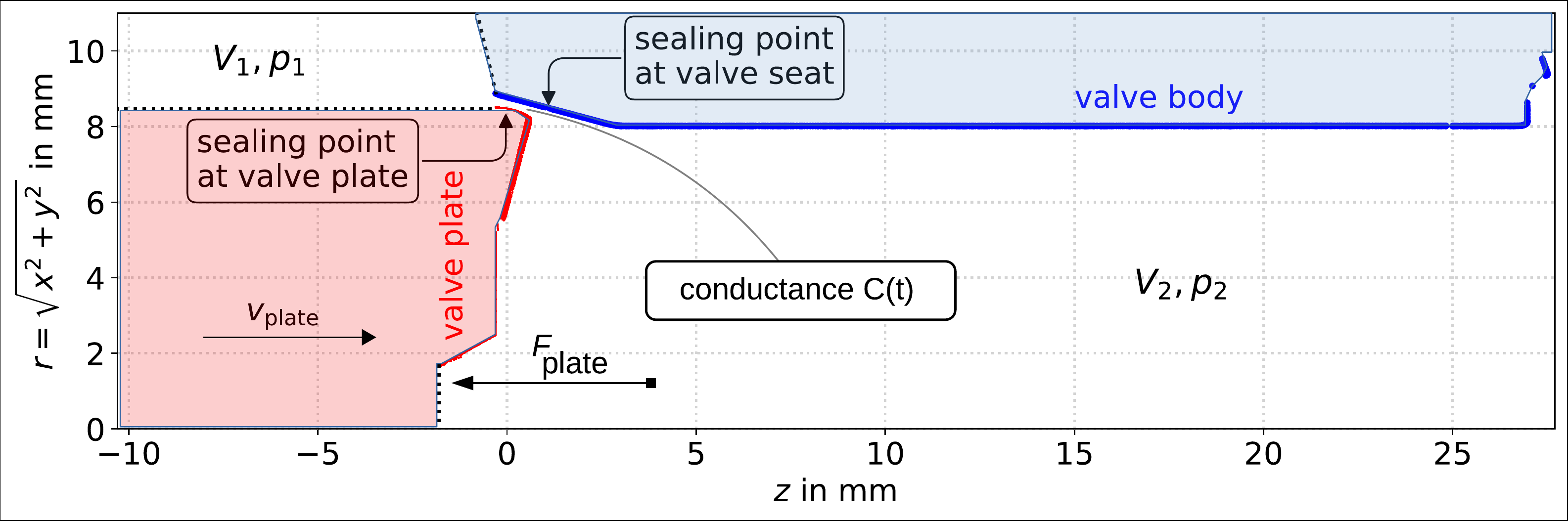}
    \caption{Scans along the inlet valve body (blue) and the valve
      plate (red). Until the valve is closed, the volumes $V_1$ and
      $V_2$ are connected via the conductance $C(t)$.  In the diagram
      the valve plate is positioned $\approx 1$\,mm in front of the
      closed position. The direction of movement and the speed of the
      valve plate during the closing process are indicated by an
      arrow.  The factor $g_\mathrm{spring}$ is equal to $1$ as long
      as there are no forces in $z$ direction.  Furthermore,
      $F_{\mathrm{plate}}$, the position of the sealing point at the
      valve plate and at the valve body is annotated.}
    \label{fig:valve_plot}
\end{figure*}

\noindent
For the purpose of the geometric determination of the contributing
volume three valves where characterized by means of a 3D coordinate
measuring machine. The dimensions of the inner volume were
measured with up to 10 $(x,y,z)$-line-scans per valve. These scans
were used to create the schema in Figure~\ref{fig:valve_plot}.

\noindent
From these measurements a model of the valve with mean coordinates and
uncertainties is derived. The sealing point of this model valve is
determined to be at a radius of $(8.46\pm0.01)$\,mm and the slope of
the valve seat is $m_\mathrm{seat}=-(0.267\pm0.002)$\,mm/mm. The
radius of the valve plate is about $8$\,mm, its length together with
its support almost $25$\,mm. The valve seat is $3$\,mm long.

\subsection{Valves closing mechanism}
\label{ssec:mech}
\noindent
Servomotors are used to open and close the valves. They provide a more
precise control of the valve plate position, velocity and closing
torque compared to a pneumatic setup. The motors are coupled to a gear
with a ratio of $R=1/169$.  The spindles of the valves have a pitch of
$s=1.4$\,mm per revolution. The linear velocity of the piston of the
valve formed by the valve plate together with the valve bellows is
given by
\begin{equation}
    \label{equ:v}
    v = \omega R s
\end{equation}
where $\omega$ is the number of revolutions per minute.  The relevant
rotational speed range lies between $\omega=20$\,rpm and
$\omega=100$\,rpm. With Equation~\ref{equ:v} one gets velocities $v$
between $2.8\times10^{-6}$\,m/s and $14\times10^{-6}$\,m/s.

The regulation circuit of the servomotor control is used to force
a constant rotation velocity. To keep $\omega$ constant,
the servomotor current $i_\mathrm{servo}$ increases with increasing torque.
The motor is stopped when the spring socket reaches its stopper and
specified torque for closing the valve is reached.

\subsection{Valve spring}
\label{ssec:spring}

\noindent
The valve's closing mechanism includes an arrangement of plate
springs. If a force $F_\mathrm{plate}$ is exerted on the valve plate,
the spring structure is deformed according to Hook's law by $\Delta
z=\frac{F_\mathrm{plate}}{D}$ where $D$ is the spring constant and
$\Delta z$ the length difference due to deformation.  The plate
springs have a spring constant $D$ of 2400\,N/mm and are pre-loaded
with a force $F_\mathrm{plate}$ of about 600\,N corresponding to a
deformation $\Delta z$ of about 0.25\,mm.  Once the valve plate
reaches the sealing point, the force on the valve plate increases.
When the force exceeds the pre-load, the deformation of the plate
springs increases according to $\Delta \dot{z}=\Delta
v=\frac{\dot{F}_\mathrm{plate}}{D}$.  Consequently, the relation
between $\omega$ and $v$ as given in Equation~\ref{equ:v} is modified.
For $\dot{F}_\mathrm{plate} > 0$, the velocity of the valve plate
$v_\mathrm{plate}=v-\Delta v$ is reduced.  This is accounted for by a
factor $g_\mathrm{spring}=1-\frac{\Delta v}{v}$\,, hence
\begin{equation}
    \label{equ:vplate}
	v_\mathrm{plate} = \omega R s g_\mathrm{spring}\,.
\end{equation}

\section{Model of gas flow during valve closing}
\label{sec:theo}

The pressure and flow conditions during the closing process can be
modeled by means of two coupled differential equations
\begin{equation}
    \label{equ:p1}
   V_1\frac{\mathrm{d}p_1}{\mathrm{d}t} =  -p_1 \frac{\mathrm{d}V_1}{\mathrm{d}t} + C(t) (p_2 - p_1)\\
\end{equation}
and
\begin{equation}
    \label{equ:p2}
   V_2\frac{\mathrm{d}p_2}{\mathrm{d}t} =  -p_2 \frac{\mathrm{d}V_2}{\mathrm{d}t} - C(t) (p_2 - p_1)
\end{equation}
based on the condition of flow conservation and the definition of flow
conductance. The temperature is assumed to be constant. Here, $C(t)$
is the time-dependent conductance formed by the gap between valve
plate and valve body. $V_1$ and $p_1$ denote the volume and the
pressure behind the valve plate (left in Figures~\ref{fig:valve_scheme}
and \ref{fig:valve_plot}). $V_2$ and $p_2$ are the starting volume
and the pressure of the  enclosed gas in front of the valve plate 
(right in the Figures~\ref{fig:valve_scheme} and \ref{fig:valve_plot}).
The pressure $p_1$ is measured by means of a Quartz Bourdon Spiral
during the initial experiments. At the later experiments $p_1$ is
measured by means of a group of 15 capacitance diaphragm gauges (SE3
group standard). $p_1$ corresponds to the filling pressure
$p_\mathrm{fill}$ introduced in Equation~\ref{equ:pgen}.

During the experiments and during a calibration the inlet valve is
closed and afterwards $p_\mathrm{fill}$ is measured. The
metrologically relevant but inaccessible quantity is the pressure $p_2
= p_\mathrm{fill} + \Delta p$.  Equation~\ref{equ:p2} can be rewritten
as
\begin{equation}
    \label{equ:dp}
    \dot p_2 = \frac{1}{V_2}\left(-p_2\dot V_2 - C(t) \Delta p(t)\right).
\end{equation}
The following sections present information on $p_2\dot V_2$ called the
\emph{volume term} and $C(t)\,\Delta p(t)$ named \emph{conductance
  term}.

\subsection{Volume term}
\label{ssec:dV}

The term $p_2\dot{V}_2$ corresponds to a flow caused by the motion of
the valve plate along the $z$ axis at a pressure $p_2$.  With an area
of $A_\mathrm{plate}=\pi r_\mathrm{plate}^2$ the volume term can be
expressed by the equation $p_2\dot{V}_2 =
p_2\frac{A_\mathrm{plate}\mathrm{d}z}{\mathrm{d}t} = p_2
A_\mathrm{plate} v g_\mathrm{spring}$. Table~\ref{tab:dv} gives an
overview for the case $g_\mathrm{spring}=1$ and a volume of
$V_2=0.02$\,l.

\begin{table}
  \centering
  \caption{Selected values of the volume term $p_2\dot{V}_2$ for
    $V_2=V_\mathrm{s}=0.02$\,l.}\label{tab:dv}
  \begin{tabular}{@{}ccc@{}}
    $\omega$ & $-\dot{V}_2$ & $-\dot{V}_2/{V_\mathrm{s}}$ \\
    in rpm & in cm$^3$/s & in 1/s \\
    \toprule
    $20$ & $6.3\times10^{-4}$ & $3\times 10^{-5}$  \\
    $80$ & $2.5\times10^{-3}$ & $1.3\times 10^{-4}$  \\
    $100$ & $3.1\times10^{-3}$ & $1.6\times 10^{-4}$ \\
    \bottomrule
  \end{tabular}
 \end{table}

The volume term increases the pressure in $V_2$.  It is independent of
the gas species and countered by the conductance term when
$\Delta p(t) > 0$.

\subsection{Conductance term}
\label{ssec:c}

The conductance $C(t)$ strongly depends on the position $z(t)$ of the
valve plate relative to the valve body.  Before entering the conical
section of the valve seat (see Figures~\ref{fig:valve_scheme} and
\ref{fig:valve_plot}) no significant pressure difference $\Delta p(t)
=p_2 - p_1$ is possible because of the large gap between valve plate
and valve body. In this range the molecular flow conductance is about
5\,l/s and the viscous conduction is even higher. Due to the conical
shape of the valve body, $C(t)$ decreases until the sealing point is
reached. For the position of the valve plate sketched in
Figure~\ref{fig:valve_plot} the gap is about $\Delta r=0.27$\,mm.

In the molecular regime, the conductance of an annular gap with inner
radius $r_\mathrm{i}$, outer radius $ r_\mathrm{o}$ and length $l$ is given by \cite{livesey}:
\begin{align}
	\label{equ:conductancemol}
	 C_\mathrm{mol}&= \sqrt{\frac{R T}{2\pi M}}\ \pi(r_\mathrm{o}^2 - r_\mathrm{i}^2)
	\Bigg(\frac{\frac{3l}{\Delta r}+\frac{l}{\Delta r+l/7}}
	 {\frac{16}{\pi}\ln(4\alpha+\frac{r_\mathrm{i}/r_\mathrm{o}+0.81}{\alpha})}
	 +1\Bigg)^{\!-1},\\\nonumber
	 \alpha&=\Big(1+\beta\tfrac{l}{2\Delta r+l}\tfrac{r_\mathrm{i}}
	 {r_\mathrm{o}}-\big(\tfrac{l}{2\Delta r+l}\tfrac{r_\mathrm{i}}{r_\mathrm{o}}\big)
	 ^{(1+\frac{r_\mathrm{i}}{r_\mathrm{o}})}\Big)^{-\frac{1}{2}}
\end{align}
where $R$ is the gas constant, $T$ the temperature of the gas, $M$ the
molar mass of the gas molecules and $\beta=0.0225$ if
$r_\mathrm{i}/r_\mathrm{o}<0.99$, otherwise
$\beta=0.225\sqrt{1-r_\mathrm{i}/r_\mathrm{o}}$. The inner radius of
the annular gap can be identified with the radius of the valve plate:
$r_\mathrm{i}=r_\mathrm{plate}$. The outer radius is given by the
radius of the valve seat that depends on the position of the valve
plate. Until the sealing point at $z=1$\,mm is reached, the outer
radius is given by
$r_\mathrm{o}=r_\mathrm{i}+m_\mathrm{seat}(1\,\mathrm{mm}-z)$.
Equation~\ref{equ:conductancemol} is valid for the molecular flow
regime. At the lowest filling pressure of $100$\,Pa, the mean free
path is $\lambda=66\,\mu$m for nitrogen corresponding to a Knudsen
number of $Kn = \lambda/d_\mathrm{h} = 1.2$ at $z=0.9$\,mm, with the
hydraulic diameter of an annular gap given as $d_\mathrm{h}=2\Delta
r$. If the valve plate continues to move along the $z$-axis, $\Delta
r$ decreases and $Kn$ increases.  On the other hand, with increasing
filling pressure $Kn$ decreases such that the assumption of molecular
flow is not valid.  In the viscous regime and laminar flow conditions,
the conductance of an annular gap is given by~\cite{wutz}:
\begin{equation}
	\label{equ:conductancevis}
	C_\mathrm{vis}= \frac{\pi}{16\eta l}
	\Big(r_\mathrm{o}^4-r_\mathrm{i}^4+
	\frac{(r_\mathrm{o}^2-r_\mathrm{i}^2)^2}{\ln(r_\mathrm{i}/r_\mathrm{o})}
	\Big)(p_1 + p_2)
\end{equation}
with $\eta$ the viscosity of the gas. Note that due to $p_1/p_2\approx
1$ no choking of the gas flow occurs. Hence, the conductance strongly
increases with increasing pressure.
The conductance over the entire pressure range is given by \cite{livesey}:
\begin{equation}
	\label{equ:conductancegap}
	C_\mathrm{gap}=C_\mathrm{vis}+
	\frac{ C_\mathrm{mol}}{1+\frac{3 \pi}{128}\frac{1}{Kn}\big (1-\big(1-
	\frac{(1 + 4 Kn)Kn^{0.45}}{1 + Kn}\big)\frac{l}{l+d_\mathrm{h}}\big)}
\end{equation}
Interestingly, the strong increase in conductance in the viscous regime only
leads to moderate reduction in the relative pressure difference of about 33\,\% for a filling
pressure of 100\,Pa and 100000\,Pa as shown in Figure\,\ref{fig:sim_ug}a. The reason
is that only close to the sealing point a significant pressure difference builds up and that
$C_\mathrm{gap}$ asymptotically approaches $C_\mathrm{mol}$ when $r_\mathrm{o}$
approaches $r_\mathrm{i}$.

\begin{table}
  \centering
  \caption{Estimation of the conductance in the molecular regime at
    room temperature and nitrogen for an annular gap with length
    $l=0.2\,$mm according to Equation~\ref{equ:conductancemol}.  $z$
    is given relative to the position shown in
    Figure~\ref{fig:valve_plot}.}\label{tab:c}
  \begin{tabular}{@{}ccc@{}}
    $z$               & $\Delta r$         & $C_{\mathrm{mol}}$  \\
    in mm           & in mm              & in cm$^3$/s           \\
    \toprule
    $0$               & $0.267$              & $1290$\\
    $0.9$            & $0.0267 $              & $48$\\
    $0.99$          & $0.00267 $              & $1.0$\\
    \bottomrule
  \end{tabular}
\end{table}

Beside the pressure dependency, $C(t)$ also depends on the gas
species. This affects $\dot{p}_2$ via Equation~\ref{equ:dp} and
therefore the amount of gas enclosed in the starting volume once the
valve is closed.

\subsection{Pressure change}
\label{ssec:pe}

\begin{figure}[t]
	\centering
        \includegraphics[angle=0, width=0.9\columnwidth]{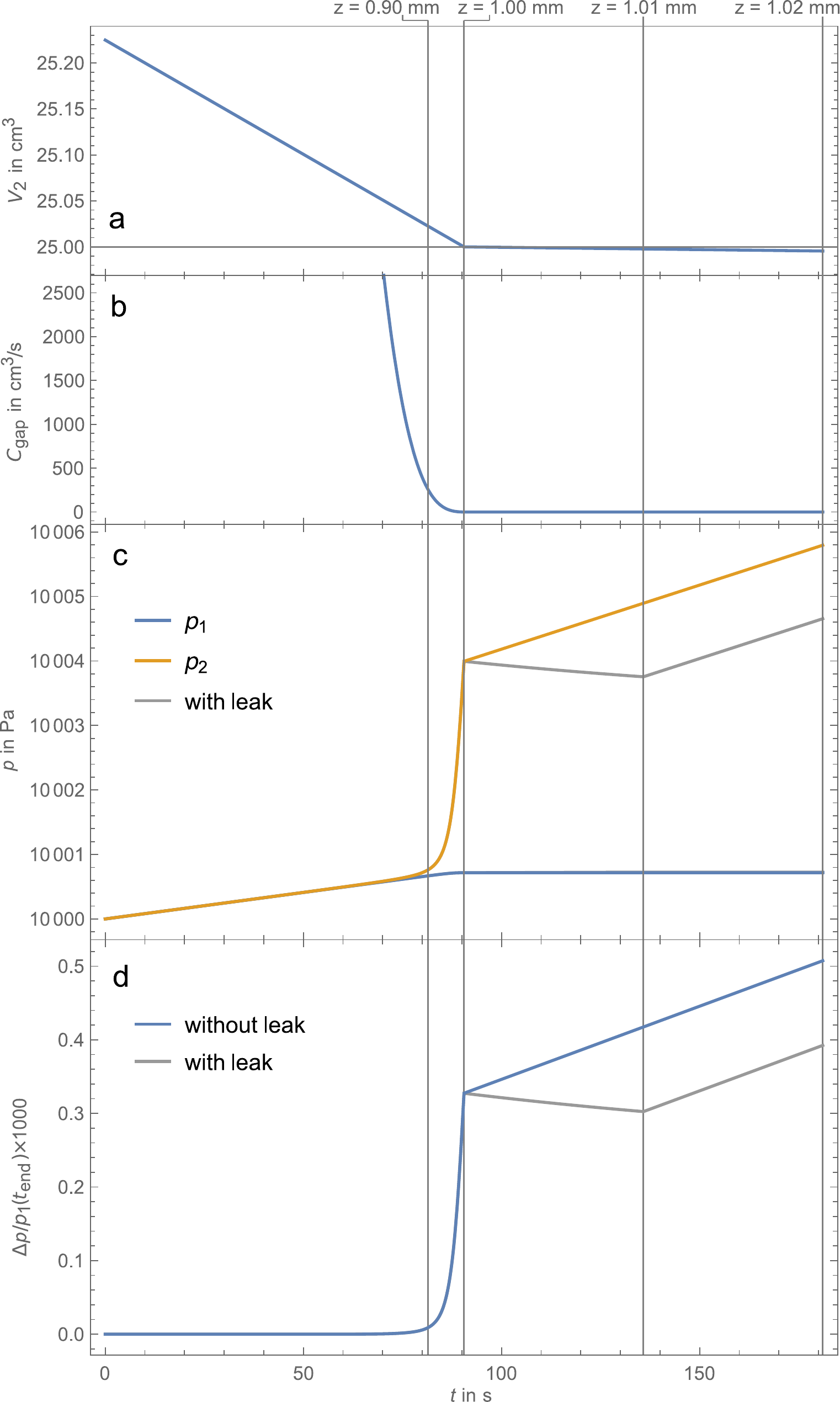}
    \caption{Calculated time dependence of the starting volume (a), flow
          conductance (b), pressures (c) and relative pressure
          difference (d) for $\omega=80$\,rpm,
          $g_\mathrm{spring}=0.02$, $V_1=3$\,l and nitrogen.  Four positions of the valve
          plate at $z=0.90$, 1.00, 1.01 and 1.02\,mm are indicated by
          vertical lines. A horizontal line in panel (a) highlights that after
          reaching $z=1.00$\,mm the change in $V_2$ is strongly reduced but not zero.}
	\label{fig:simulation}
\end{figure}

The time-dependent pressure change can be obtained by numerical
solution of Equations~\ref{equ:p1} and~\ref{equ:p2} with $C(t)$ from
Equations~\ref{equ:conductancemol} to \ref{equ:conductancegap}. The change in
volume is solely attributed to $V_2$, and $\dot{V}_1$ is set to zero.
Other choices are possible but will not affect the result because a
significant pressure difference only builds up shortly before the
closing position of the valve plate is reached.

Figure~\ref{fig:simulation} shows the calculated time-dependent change of the
starting volume (a), flow conductance (b), pressures (c) and relative
pressure difference (d) for $\omega=80$\,rpm, $g_\mathrm{spring}=0.02$,
$V_1=3$\,l and nitrogen. Four positions of the valve plate at
$z=0.90$, 1.00, 1.01 and 1.02\,mm are indicated by vertical lines. The
first line marks the position at which a significant pressure
difference builds up. The second line marks the position at which the
valve plate starts to seal against the valve seat and the velocity of
the valve plate is strongly reduced to $v g_\mathrm{spring}$.  The
third line marks the positions up to which a potential leak after
sealing against the valve seat persists.  The last line marks the
position at which the specified closing torque of the valve is reached
and the movement of the valve plate is stopped.

In Figure~\ref{fig:simulation}c the pressure in the starting volume
$p_2$ (yellow) and in the volume of the gas inlet $p_1$ (blue) are
plotted as a function of time together with the pressure change for
a situation in which a leak of $0.2$\,cm$^3/$s persists up to
$z=1.01$\,mm.  Four different phases of the valve-closing process can
be identified. Up to about $z=0.90$\,mm, the change in volume is
constant and causes a relative change in pressure according to
$\dot{p}_1/p_1=\dot{p}_2/p_2=-\dot{V}_2/(V_1+V_2)$. The flow
conductance is decreasing but no significant pressure difference is
generated down to about $300$\,cm$^3/$s (see
Figure~\ref{fig:simulation}b and~\ref{fig:simulation}d). Below this
conductance, the pressure in the staring volume rises strongly
approaching $\dot{p}_2/p_2=-\dot{V}_2/V_2$. When the valve plate
reaches the sealing point ($z=1.00$\,mm) its movement is strongly
reduced because of the force exerted.  The change in pressure in the
starting volume is now
$\dot{p}_2/p_2=-g_\mathrm{spring}\dot{V}_2/V_2$. The increase in
pressure maybe partly counteracted by a gas flow through a leak until
the valve is fully sealed at $z=1.01$\,mm (gray lines in
Figure~\ref{fig:simulation}c and d). The pressure in the starting volume
continues to increase until the closing torque is reached and the
movement of the valve plate is stopped ($z=1.02$\,mm). The model
contains four valve-specific parameters that must be adjusted to match
the measurements: the reduction in movement of the valve plate
$g_\mathrm{spring}$ once the sealing point is reached, the position at
which the closing torque is reached, the size of a potential leak, and
the length of the channel of the annular conductance formed by the
valve plate and the seat. The latter significantly impacts the
pressure difference that is built up in the moment the sealing
position is reached and is chosen to be $0.2$\,mm here.

\section{Experiment}
\label{sec:experiment}

The differential pressures were investigated systematically by varying
the parameters $V_1$, $V_2$, $\omega$, $p_1$ and the gas species.
\begin{figure}[t]
    \centering
    \includegraphics[angle=0, width=\columnwidth]{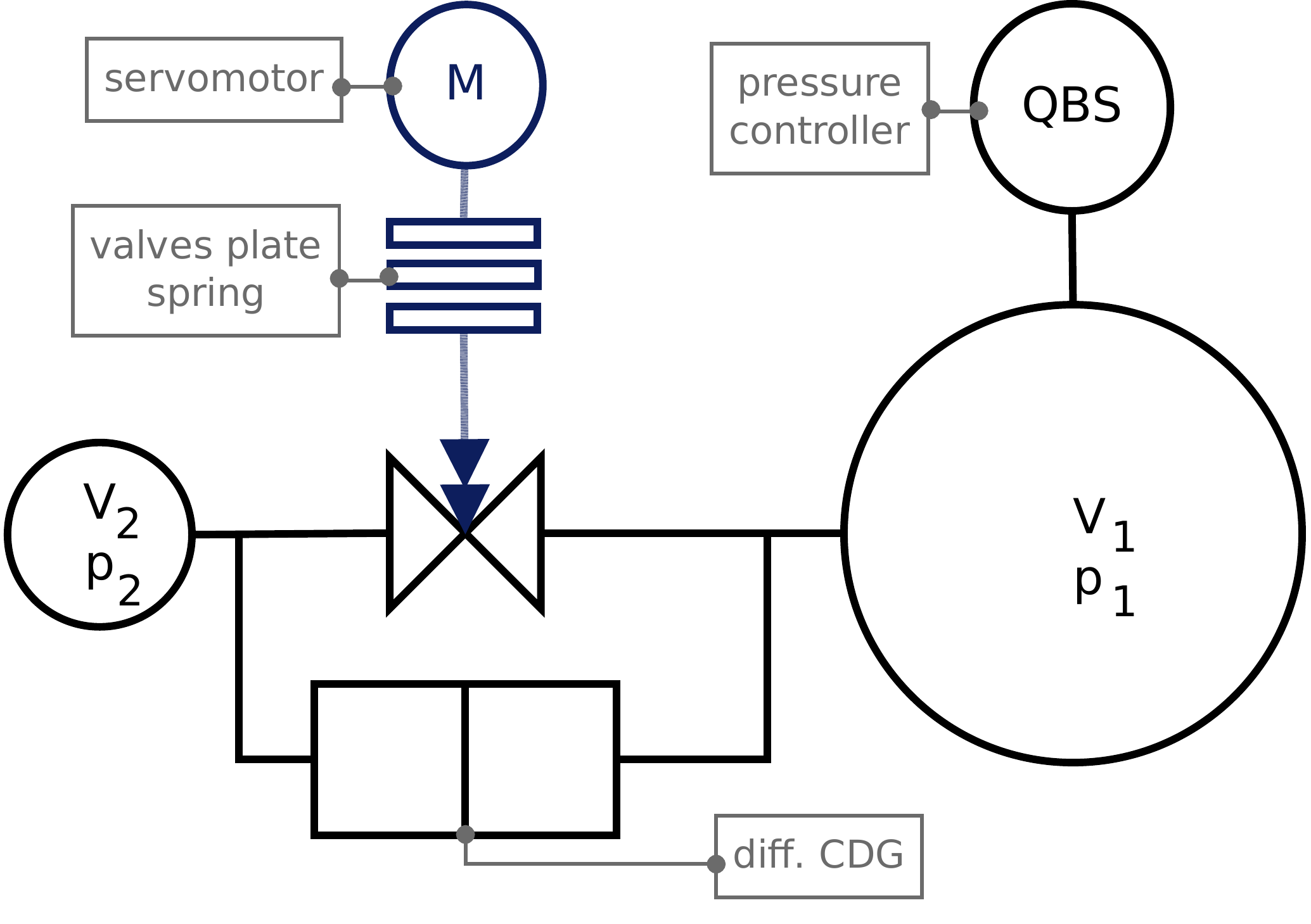}
    \caption{Experimental setup. A Quartz Bourdon Spiral (QBS) was
      used to set and measure the pressure $p$. During the valve
      closing process the differential pressure $\Delta p(t)$ across
      the valve plate as well as the servomotor current
      $i_\mathrm{servo}$ are recorded. The experiment was realized
      with two different valves of the same model.}
    \label{fig:exp_scheme}
\end{figure}
Figure~\ref{fig:exp_scheme} shows a sketch of the experimental setup.
$\Delta p(t)$ was measured by means of a differential capacitance
diaphragm gauge with a full scale of $1.3$\,kPa. The heater of the CDG
was turned off.  For each measurement run, the valve plate was set
about 2\,mm in front of the closing position, where the offset
corrected signal of the CDG is still zero.  Then the closing process
was initiated and the CDG measurement signal was recorded as a
function of time until the closing torque was reached. In some cases
the current consumption of the servomotor was recorded
simultaneously. The measurements were repeated several times in order
to examine the repeatability.

\begin{figure}[]
    \centering
    \includegraphics[angle=0, width=\columnwidth]{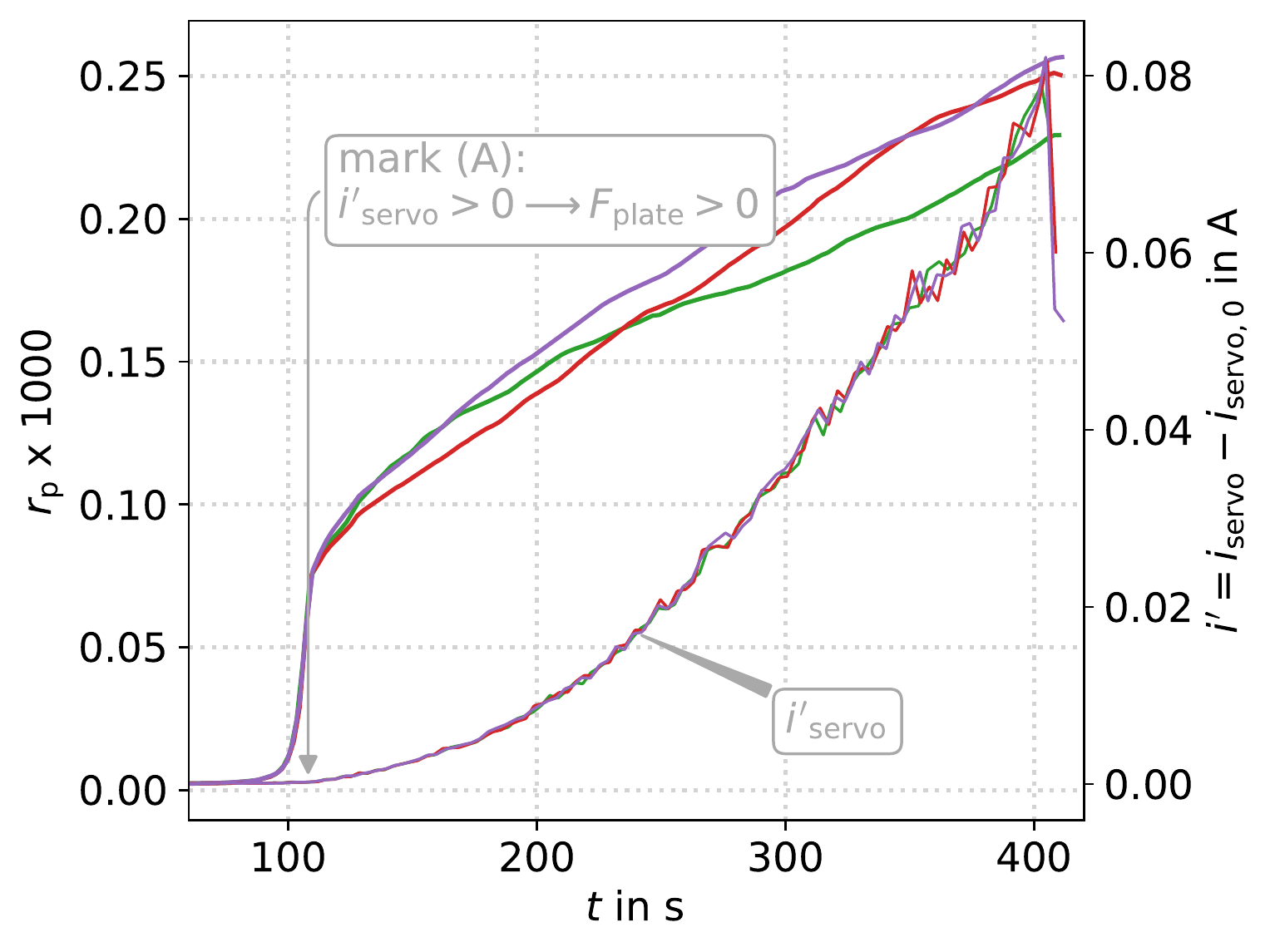}
    \caption{Three repeated measurements of the servomotor current
      $i'$ (axis on the right) and the relative pressure difference
      $r_p$ (axis on the left) at $\omega=20$\,rpm, $V_1=3$\,l and a
      volume $V_2=25$\,ml. The ripples in the current tracks may point
      to a discontinuous movement of the valve plate.}
    \label{fig:current}
\end{figure}

All measurements show more or less the characteristics plotted in
Figure~\ref{fig:current}. The graph can be roughly separated in two
parts. Up to the position marked with (A), the slope of the relative
pressure change $r_p=\frac{\Delta p(t)}{p_1}$ depends on the volume
$V_2$. Beside the closing speed dependency the height of $r_p$
corresponds to the travel length, the valve plate can move without an
occurring $F_{\mathrm{plate}}$ in the region with low conductivity.
After the position marked with (A), the slope of $r_p$ depends mainly
on the valve spring constant $D$.

Figure~\ref{fig:current} shows $r_p$ and the current
$i'=i_{\mathrm{servo}}-i_{\mathrm{servo},0}$ for three repeated
measurements.  $i_{\mathrm{servo}}$ denotes the total current
consumption of the servomotor and $i_{\mathrm{servo},0}$ the current
for $F_{\mathrm{plate}} = 0$. $r_p$ and $i_{\mathrm{servo}}$ are
simultaneously recorded. At mark (A) $i'$ rises significantly. This
can be related to the increase of the force on the valve plate by
friction and an associated torque $\tau_\mathrm{servo}$:
$F_\mathrm{plate}\propto \tau_\mathrm{servo}\propto i'$.  The spring
deforms and the velocity of the valve plate is strongly reduced (see
section~\ref{ssec:spring}). The contribution of the volume term is
reduced by the same factor which results in a slope with a smaller
value in the $\Delta p(t)/p_1$ vs. $t$ chart.


\subsection{Volume variation}

\begin{figure}
    \centering
    \includegraphics[angle=0, width=\columnwidth]{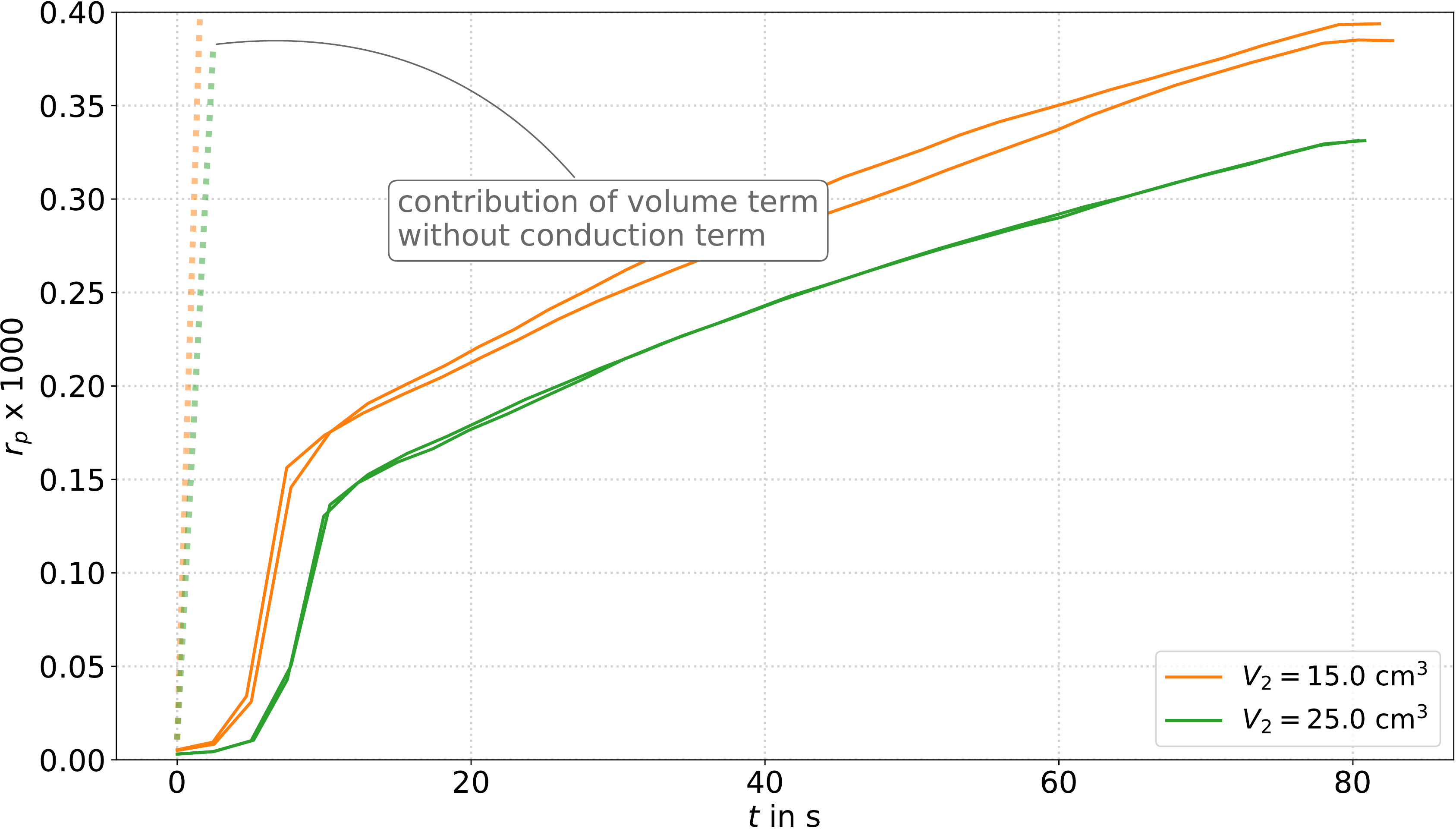}
    \caption{Relative pressure difference as a function of time for
      two different volumes $V_2$ with $p_1=100$\,kPa, $V_1=3$\,l and
      $\omega=100$\,rpm}
    \label{fig:var_vol}
\end{figure}

Equation~\ref{equ:dp} shows that $\dot p_2 \propto
\frac{1}{V_2}$. This dependency is confirmed by the experimental
results shown in Figure~\ref{fig:var_vol}.  Both series of
measurements with three repetitions show the expected dependency in
the slope of the resulting $r_p$. The diagram includes the theoretical
values for the contribution of the \emph{volume term} for the case
$V_2 = 15$\,ml and $V_2 = 25$\,ml without the \emph{conductance
  term}. Due to the omitted $C(t)\,\Delta p(t)$ part the theoretical
slopes are larger compared to the measured slopes.

The measurements shown in Figure~\ref{fig:var_pressure} aimed also on
the examination of the influence of $V_1$ on $p_1$. $V_1$ has no
measurable influence on $\Delta p(t)$ at least in the case $V_1 \gg V_2$.

\subsection{Closing speed variation}

\begin{figure}
    \centering
    \includegraphics[angle=0, width=\columnwidth]{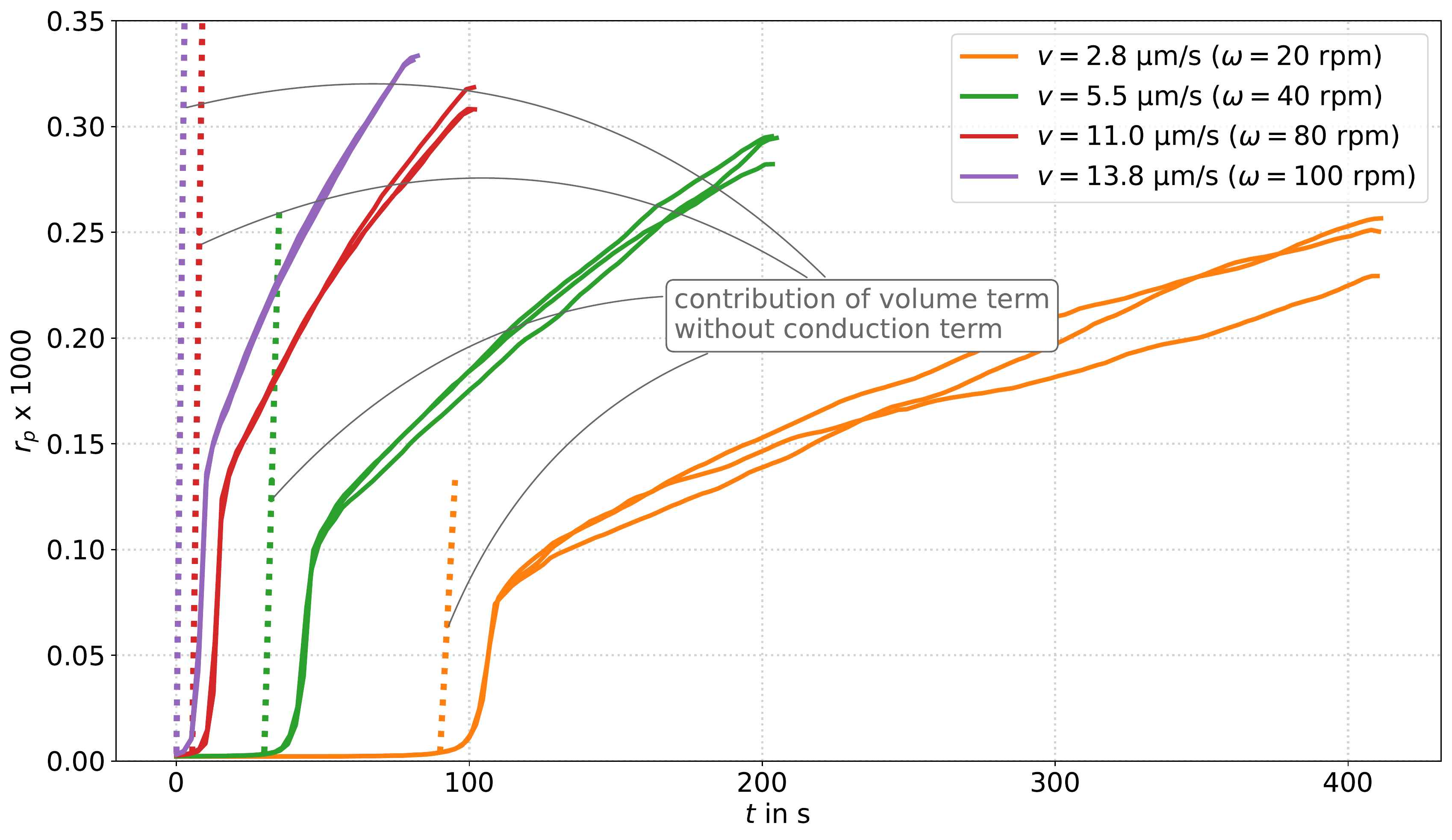}
    \caption{Relative pressure difference as a function of time for
      different closing speeds with $p_1=10$\,kPa, $V_1=3$\,l and
      $V_2=25$\,ml}
    \label{fig:var_speed}
\end{figure}

The influence of the closing speed $\omega$ was examined by the
measurements plotted in Figure~\ref{fig:var_speed}. The conductance
term causes a back flow $\Delta p_\mathrm{bf}(t)
=\int_{t_0}^{t}\,C(t')\,\Delta p(t')\,\mathrm{d}t'$. This means that
the longer the closing process lasts the more the pressure rise in
$V_2$ is reduced by the conductance term. Like in
Figure~\ref{fig:var_vol} the theoretical values for the contribution
of the \emph{volume term} for the case $V_2 = 25$\,ml without the
$C(t)\,\Delta p(t)$ part are included.

\subsection{Influence of valve tolerances}

The data shown in Figure~\ref{fig:var_pressure} were measured using
another valve of the same type as the one used for
Figures~\ref{fig:var_vol} and~\ref{fig:var_speed}.  A different shape
and height of the $r_p$ graph can be observed.  Especially the
constant or partially decreasing pressure in the range $30$\,s to
$60$\,s is remarkable. An explanation may be a slowed down movement
of the valve plate along the z-axis or a leak that persists until the
valve is fully sealed as depicted in Figure~\ref{fig:simulation}d.
During the period when the movement stops or slows down, the
conductance term lowers the $\Delta p(t)$.

\subsection{Pressure variation}
\label{ssec:var_p}
\begin{figure}
    \centering
    \includegraphics[angle=0, width=\columnwidth]{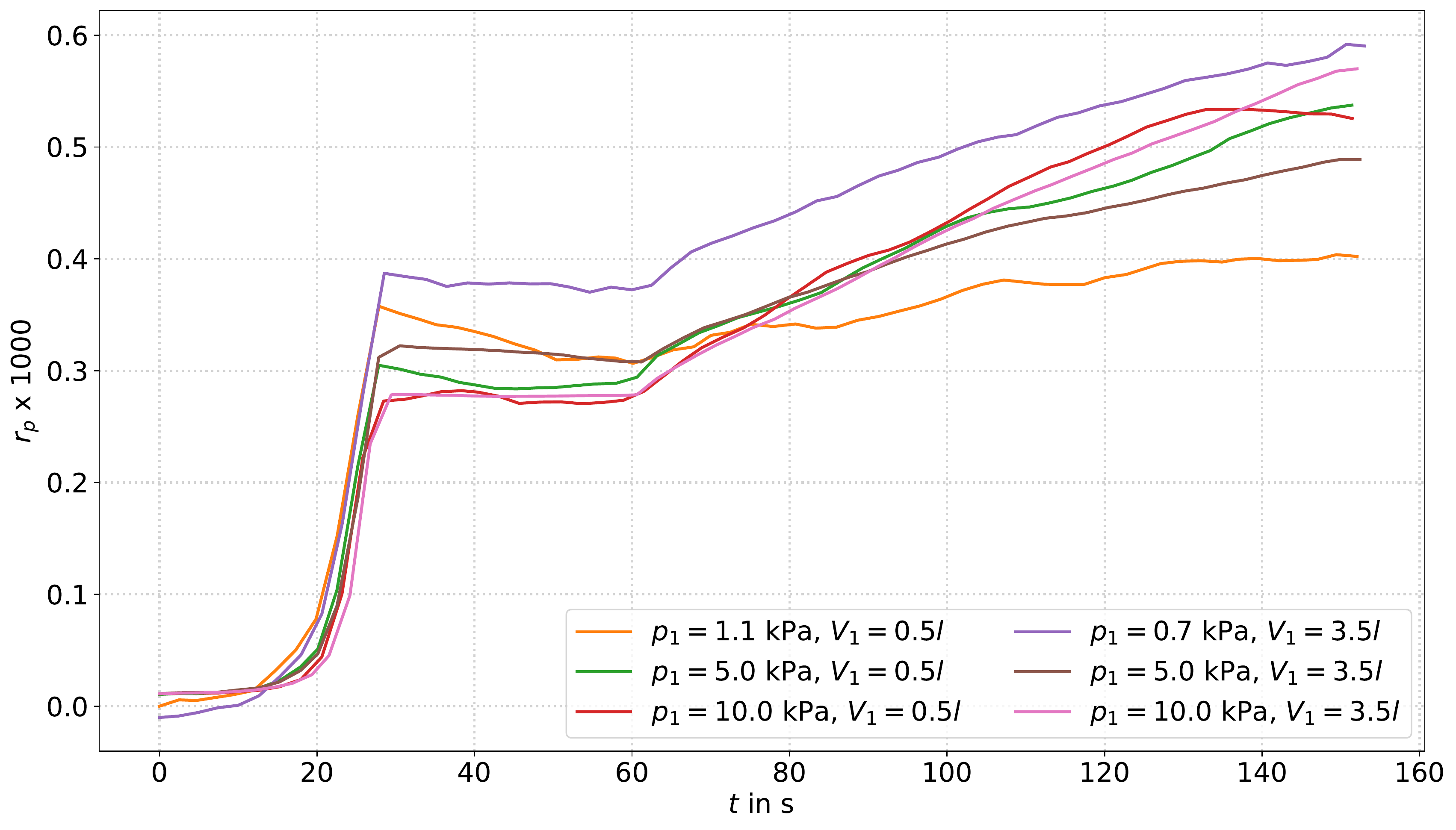}
    \caption{Relative pressure difference as a function of time for different
    	volumes $V_1$ and pressures $p_1$ with $V_2=25$\,ml and $\omega=80$\,rpm}
    \label{fig:var_pressure}
\end{figure}

The influence of $p_1$ on $\Delta p(t)$ in
Figure~\ref{fig:var_pressure} is expected from the considerations
given in section~\ref{ssec:c}: in the viscous flow regime an
increasing pressure causes an increasing conductance. With a larger
$C$ the amount of back flow from $V_2$ to $V_1$ becomes bigger and the
pressure increase in $V_2$ is reduced more strongly (see
Equation~\ref{equ:dp}).

\subsection{Influence of the gas species on the expansion ratio}
\label{sec:gasspeciesexpansion}

$C$ depends on the gas species. In the molecular regime
$C\propto 1/\sqrt{M}$ where $M$ is the molar mass. Under laminar flow
conditions $C \propto 1/\eta$ where $\eta$ is the viscosity
of the gas.  The effective expansion ratio of SE3 starting from $V_\mathrm{s}$ was determined for
different gas species. $f_\mathrm{eff,s}$ is determined from the pressure ratio:
\begin{equation}
    \label{equ:p/p}
    f_\mathrm{eff} = \frac{p'_\mathrm{after}}{p'_\mathrm{fill}}\frac{T'_\mathrm{before}}{T'_\mathrm{after}}\,.
\end{equation}
as shown in~\cite{frs5}.  In Equation~\ref{equ:p/p}
$p'_\mathrm{fill}$, $T'_\mathrm{before}$, $p'_\mathrm{after}$ and
$T'_\mathrm{after}$ are the pressures and temperatures before and
after the expansion at the time of determining $f_\mathrm{eff}$.
$p'_\mathrm{after}$ was measured by means of a large area non-rotating piston gauge of the type 
FRS5~\cite{frs5} and a differential CDG. $p'_\mathrm{fill}$ was
obtained by using the SE3 group standard.

For this method the following has to be considered: since
$p'_\mathrm{fill}$ was measured after the closing of the inlet valve
of the starting volume $p'_\mathrm{before}=p'_\mathrm{fill}+\Delta p'$.
Thus, $\Delta p'$ enlarges
$p'_\mathrm{after}$ by $2\times10^{-4}$ to $6\times10^{-4}$ as the
results of the previous section have shown.  The value of the
effective expansion ratio determined from a pressure ratio contains
this effect: the enlarged $p'_\mathrm{after}$ causes an enlarged
$f_\mathrm{eff,s}$. The subscript ``$\mathrm{eff}$'' reflects the
determination of the effective expansion ratio from a pressure ratio
in contrast to the determination from a volume ratio that is referred
to as $f$.

The largest effect of the gas species on $f_\mathrm{eff}$ is expected
for the smallest volume $V_\mathrm{s}$, since the volume $V_2$ (see
Equation~\ref{equ:p2}) has to be small to gain a measurable effect.
In order to use FRS5 for the measurement of $p'_\mathrm{after}$ with
low uncertainties, the gas inside the calibration vessel has to be
accumulated~\cite{nplreport, redgrave99}. With a filling pressure of
$100$\,kPa and a nominal $f_\mathrm{eff,s}$ of $1\times10^{-4}$ one
expansion step generates $10$\,Pa. The results of the investigation of
the gas species dependency of $f_\mathrm{eff,s}$ are given in
Table~\ref{tab:exp_gas_dep}.

\begin{table*}
  \centering
  \caption{Result of expansion ratio measurements carried out
    according the method described in~\cite{nplreport, redgrave99}
    and~\cite{frs5} for different gas species. Given are the
    proportional factors $1/\sqrt{M}$ and $1/\eta $ of the gas
    species. Furthermore, the determined values of $f_\mathrm{eff,s}$ and the
    expanded relative type A measurement uncertainties with a coverage
    factor of $k=2$.}\label{tab:exp_gas_dep}
    
  \begin{tabular}{@{}rcccc@{}}
    gas      & $1/\sqrt{M}$          & $1/\eta$              & $f_\mathrm{eff,s}$                 & $u_a(f_\mathrm{eff,s})$ \\
             & g$^{-1/2}$ mol$^{1/2}$  & $\mu$Pa$^{-1}$ s$^{-1}$ &                       &    \\
    \toprule
    helium   & $5.0\times 10^{-1}$   &$5.0\times 10^{-2}$      &$1.05723\times 10^{-4}$ & $7.2\times 10^{-8}$ \\
    nitrogen & $1.9\times 10^{-1}$   &$5.6\times 10^{-2}$      &$1.05759\times 10^{-4}$ & $5.8\times 10^{-8}$ \\
    argon    & $1.6\times 10^{-1}$   &$4.4\times 10^{-2}$      &$1.05739\times 10^{-4}$ & $3.7\times 10^{-8}$ \\
  \bottomrule
  \end{tabular}
\end{table*}

With regard to the expansion ratios $f_\mathrm{eff,s}$ measured at SE3, the
differences between the values are not significant due to the
measurement uncertainties.

\subsection{Influence of closing time on calibration results}

With the modified pneumatic valves of the type
\emph{VAT~F57--74423--02} mentioned in section~\ref{sec:ini} the
experiments described in the following are carried out. The effective
accommodation coefficient $\sigma = p_\mathrm{ind}/p_\mathrm{after}$ of
two spinning rotor gauges (SRG) \cite{fremerey85} are determined at a
pressure of $3\times10^{-2}$\,Pa. Here $p_\mathrm{ind}$ denotes the
pressure indicated by the SRG at the pressure
$p_\mathrm{after}$ generated by the primary standard SE1. The starting
volume has a nominal value of $17$\,ml. The calibration vessel has a
size of $230$\,l, such that the filling pressure was approximately
$400$\,Pa. The corresponding expansion ratio $f_\mathrm{eff,1}$ was determined via
a pressure ratio in the same way as described in the previous
section. For this determination the starting volume's inlet valve was
operated with a closing time of $t_\mathrm{close} = 30$\,s. This
corresponds to a closing speed of $67$\,$\mu$m/s.

\begin{figure}
    \centering
    \includegraphics[angle=0, width=\columnwidth]{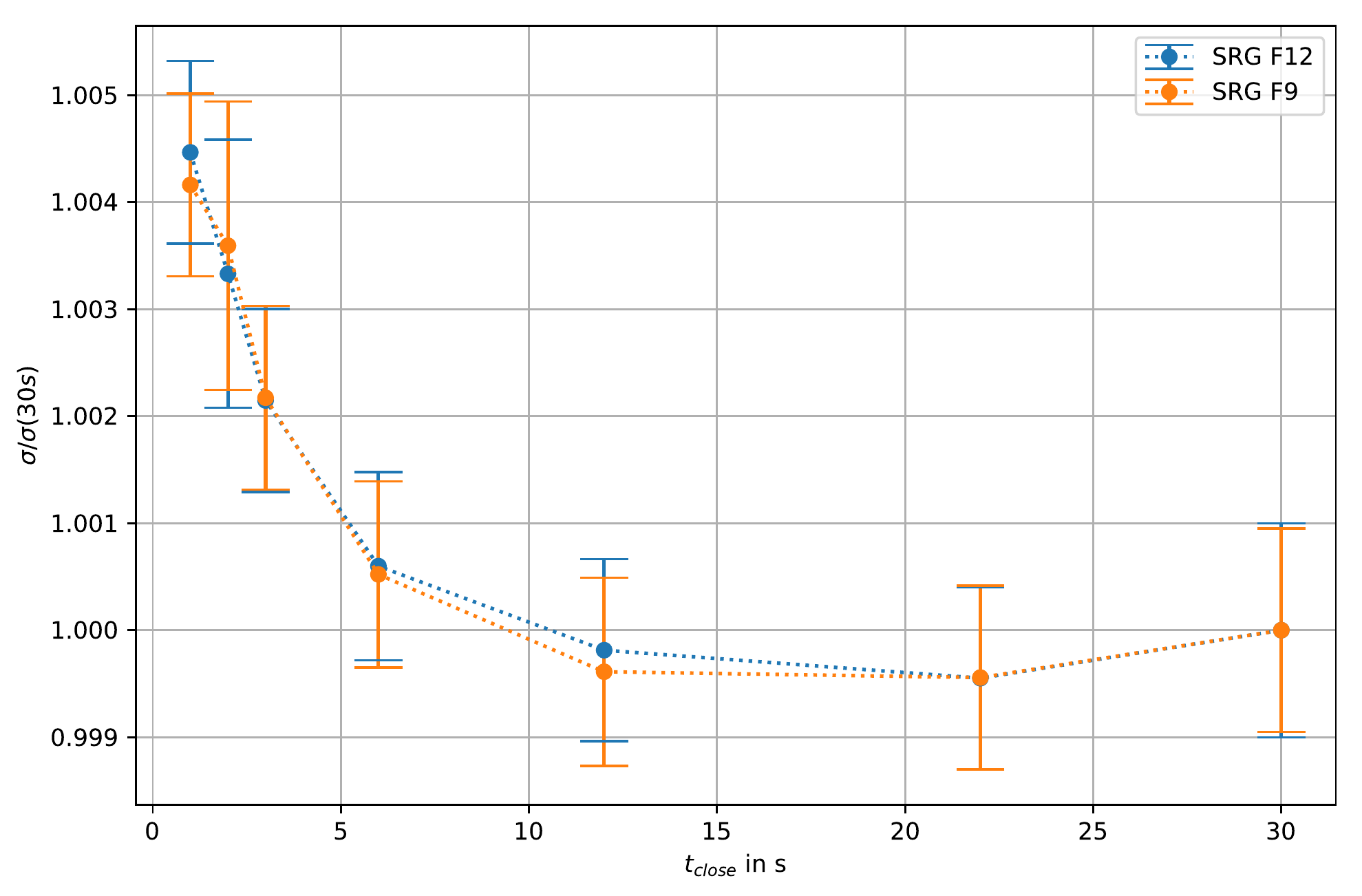}
    \caption{Influence of the variation of the closing time on
      calibration results of the static expansion system SE1. With the
      help of a throttle in the pneumatic system the closing time
      $t_\mathrm{close}$ was varied between $1$\,s and $30$\,s. The
      effective accommodation coefficients $\sigma$ of the two SRGs
      (F9 and F12) are normalized to $\sigma(\mathrm{30s})$ since
      the used expansion ratio $f_\mathrm{eff,1}$ was determined with
      $t_\mathrm{close} = 30$\,s. The error bars indicate the combined
      Type~A and Type~B standard measurement uncertainty.}
    \label{fig:var_t_close}
\end{figure}

Figure~\ref{fig:var_t_close} shows the calibration results for a
successive reduction in closing time. The corresponding closing speeds
vary between $2$\,mm/s and $67$\,$\mu$m/s. $t_\mathrm{close}$ was
changed by means of the throttle in the pneumatic system of the inlet
valve. The results of Figure~\ref{fig:var_t_close} show that
differences up to $0.4$\,\% occur when the valve is closed with a
speed of $2$\,mm/s compared to the value obtained with the speed used
for the determination of $f_\mathrm{eff,1}$. These deviations are
caused by the enlarged calibration pressure due to the dependence of
$r_p$ on the closing speed.

\section{Discussion}

\subsection{Temperature influence}
\label{ssec:dT}

The experiments described in section~\ref{sec:experiment} were carried
out in a room where the temperature changes are less than
$0.15$\,K/h. This maximum temperature change results in a relative
change in pressure of $8\times10^{-6}$\,min$^{-1}$. These variations
have a random characteristics. A potential heat contribution caused by
friction between the plate and the valve seat is considered to be
negligibly small.

\subsection{Influence of valve bellows volume change}

The change in the volume of the valve due to deformation of its
bellows (see Figure~\ref{fig:valve_scheme}) $\Delta V_\mathrm{bellow}$
is negligible too: on the one hand $\Delta V_\mathrm{bellow}$ relates
to the volume $V_1 + V_2$ until $V_1$ and $V_2$ are separated by a low
$C$. When $V_1$ and $V_2$ are separated $\Delta V_\mathrm{bellow} \ll
V_1$ applies safely. At least, Figure~\ref{fig:var_pressure} shows no
indication of a significant influence of $\Delta V_\mathrm{bellow}$.

\subsection{Discontinuous closing movement}

Figure~\ref{fig:valve_plot} suggests that there is no contact between
the valve plate and the seat until closing torque is reached.
However, if the valve is open, the valve plate can be moved freely by
about $1$\,mm to $2$\,mm perpendicular to the $z$ axis. This is an
intended engineering behavior since it is necessary to avoid a
mechanical over determination. The observed flattening of the $r_p$
curve in Figure~\ref{fig:var_pressure} can be explained by a
discontinuous movement in combination with a small conductance
$C$. Moreover, Figure~\ref{fig:current} gives indications for a stick
slip motion by the $i'_\mathrm{servo}$ signal especially for small
$\omega$. Since the stick slip friction has a random characteristic,
the overall final $\Delta p(t)$ contains some random contributions.

\section{Implications for the generated calibration pressure}

In a static expansion system, the generated pressure is obtained
from the filling pressure and the expansion ratio according to
Equation~\ref{equ:pgen}.
If the expansion ratio $f$ is determined from the absolute values of
the volumes $V_\mathrm{s}$ and $V_\mathrm{e}$, the filling pressure
must be corrected by the factor
$\widetilde{K}=(1+\frac{\Delta p}{p_\mathrm{fill}})=(1+r_p)$.
In such a situation it is advisable to ensure that $r_p$ is minimal,
e.\,g. by reducing the speed of the valve.

Alternatively, an effective expansion ratio $f_\mathrm{eff}$ can be determined from a
measurement of a filling pressure $p'_\mathrm{fill}$ and the
pressure after expansion $p'_\mathrm{after}$ as described in
section~\ref{sec:gasspeciesexpansion}. From Equation~\ref{equ:p/p}
and \ref{equ:pgen} it follows:
\begin{align}
\label{equ:feff}
&f_\mathrm{eff}=\Big(1+\frac{\Delta p'}{p'_\mathrm{fill}}\Big)f
\end{align}
When using this method the effect of $\Delta p$ is calibrated into
$f_\mathrm{eff}$ for the conditions used during its determination.
Substituting Equation~\ref{equ:feff} into Equation~\ref{equ:pgen},
the generated pressure is given by
\begin{equation}
p_\mathrm{after}
=p_\mathrm{fill}\frac{1+\frac{\Delta p}{p_\mathrm{fill}}}{1+\frac{\Delta p'}{p'_\mathrm{fill}}}
f_\mathrm{eff}\frac{T_\mathrm{after}}{T_\mathrm{before}}
\end{equation}
and the correction factor becomes $K=(1+r_p)/(1+r'_p)$. Thus, $K=1$
when the conditions during use are
the same as during the determination of the effective expansion ratio.

The considerations in section~\ref{sec:theo} showed that a pressure
and gas species dependence of $r_p$ can be expected. Both dependencies
are introduced by the conductance term. The closing process takes
place within the viscous flow regime. Here a smaller
$p_{\mathrm{fill}}$ causes a smaller $C$. A smaller molecular mass
causes a larger $C$ with a $1/\sqrt{M}$ and a smaller viscosity causes
a larger $C$ with $1/\eta$. Due to the high measurement uncertainty
and the small size of the effect, it was not measurable as
Table~\ref{tab:exp_gas_dep} shows. However, a correction factor should
be introduced.

\subsection{Pressure dependent correction factor}

\begin{figure}
    \centering
    \includegraphics[angle=0, width=\columnwidth]{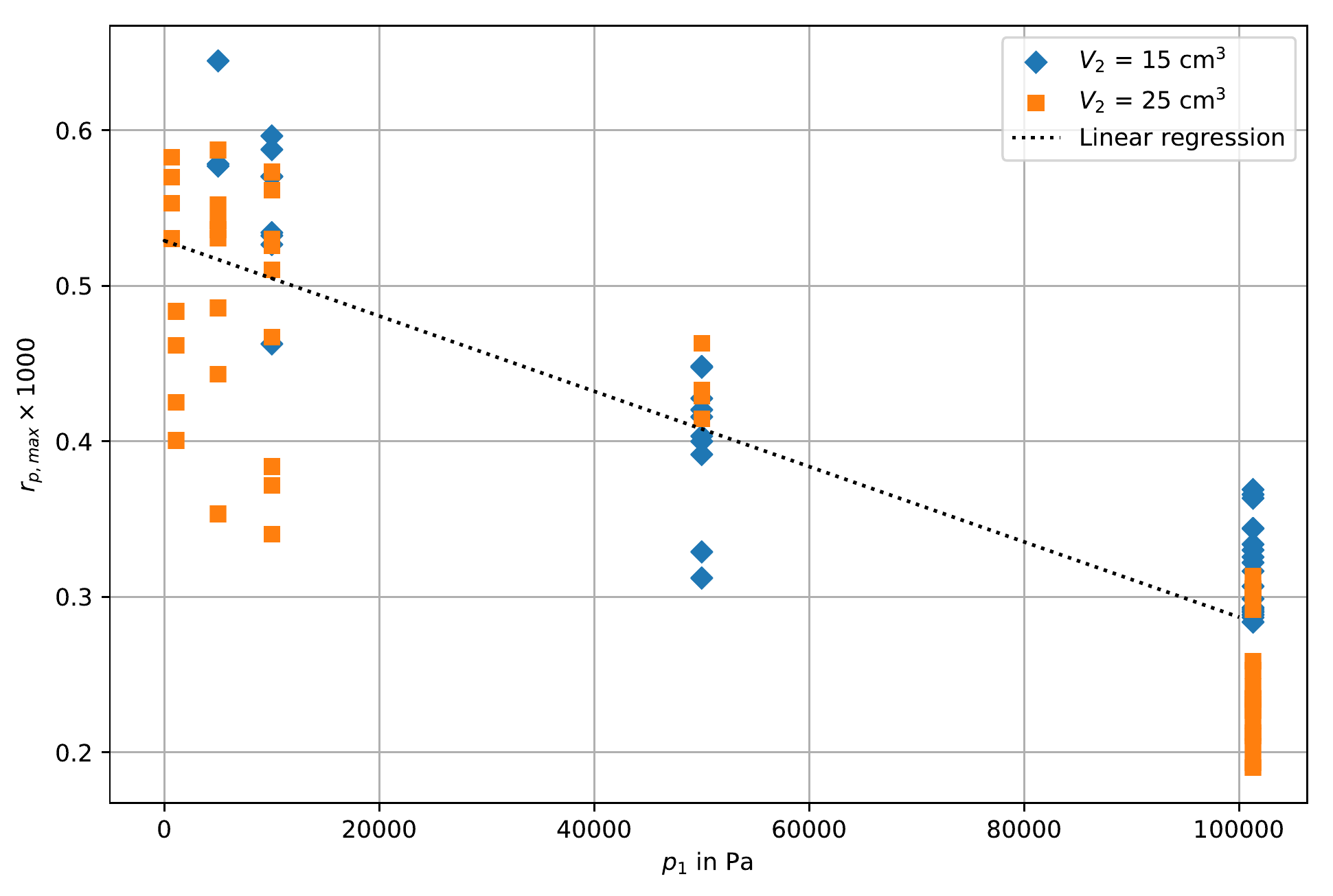}
    \caption{Experimental data for various $V_1$, a closing speed of
      $\omega=80$\,rpm, $V_2=0.015$\,l and $V_2=0.025$\,l together with a
      linear regression of the data.}
    \label{fig:p_corr}
\end{figure}

For the calibration pressures generated by means of $V_\mathrm{s} = 0.02$\,l a
correction $K_p$ is derived on the basis of the experimental data
shown in Figure~\ref{fig:p_corr}. Here, the maximum reached $r_p$, measured at a closing speed of
$\omega=80$\,rpm are plotted across the pressure $p_\mathrm{fill}$ for the volumes $V_2=0.015$\,l and $V_2=0.025$\,l. $V_1$ lies
between $0.5$\,l and $3.5$\,l. From these data the slope $\widetilde{m}$ and
offset $b$ are derived by means of a linear regression:
$r_p(p_\mathrm{fill})=\widetilde{m}\,p_\mathrm{fill} + b$. The
correction factor is then given by
\begin{align}
	K_p &= \frac{1 + \widetilde{m}\,p_{\mathrm{fill}} + b}{1 + \widetilde{m}\,p'_{\mathrm{fill}} + b} =
	1 + m\,(p_{\mathrm{fill}}-p'_{\mathrm{fill}})\nonumber\\
	&= 1 - 2.4\times10^{-9}\,\textrm{Pa}^{-1} (p_{\mathrm{fill}} - 101300\,\mathrm{Pa})
\end{align}
where $m=\widetilde{m}/(1 + \widetilde{m}\,p'_{\mathrm{fill}} + b)$.
At $101.3$\,kPa, the filling pressure used to determine $f_\mathrm{eff,s}$, $K_p-1$
becomes $0$. For $100$\,Pa, the smallest $p_{\mathrm{fill}}$, $K_p$
increase the calibration pressure by a relative value of
$2\times10^{-4}$.  $K_p$ has a relative standard measurement
uncertainty of $2.3\times10^{-5}$.

\subsection{Measurement uncertainty due to gas-species dependency}

\begin{figure}
	\centering
	\includegraphics[angle=0, width=\columnwidth]{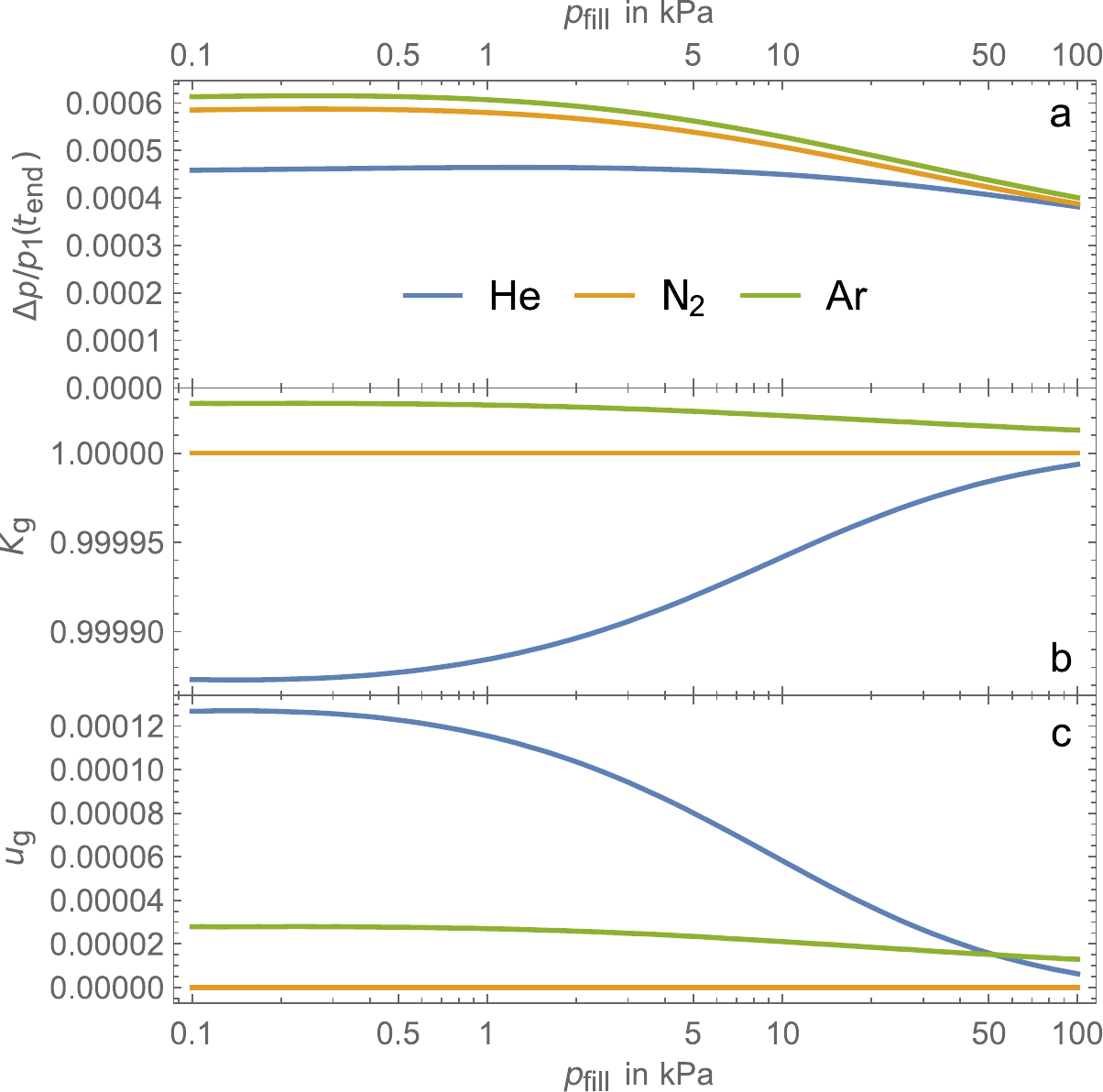}
	\caption{Relative pressure difference (a), correction factor (b) and
		standard uncertainty (c) for helium and argon relative to
		nitrogen as a function of filling pressure. The values are
		calculated for a valve without leak (cf. Fig.\,\ref{fig:simulation})
		and a closing speed of 80\,rpm.}
	\label{fig:sim_ug}
\end{figure}

Since the differences between the values of $f_\mathrm{eff,s}$ for nitrogen, argon
and helium are not significant (see Table~\ref{tab:exp_gas_dep}),
the dependency of $f_\mathrm{eff,s}$ on the gas species will not be corrected.
Instead it will be accounted for by an additional gas species and filling
pressure dependent uncertainty contribution based on the numerical model
introduced in section~\ref{sec:theo}. In Figure~\ref{fig:sim_ug} the
relative pressure difference (a), the correction factor (b) and
standard uncertainty (c) for helium and argon relative to
nitrogen are shown as a function of filling pressure. The values are
calculated for a valve without leak (cf. Fig.\,\ref{fig:simulation})
and a closing speed of 80\,rpm. The correction factor is calculated
according to $K_g=(1+r_p(p_\mathrm{fill},g))/(1+r_p(p_\mathrm{fill},\mathrm{N}_2))$
with $g=$ He, Ar. The contribution to the standard measurement uncertainty
is estimated by $u_g=|K_g-1|$.

\section{Summary and Conclusions}

Due to the valve closing process pressure differences occur depending
on the volume, closing speed, filling pressures and the gas
species. The effects can be described by a differential equation
derived from the condition of flow conservation. For a DN16CF angle valve and volumes
of a size around $0.02$\,l (size of the smallest starting volume of
SE3) relative pressure differences of some $10^{-4}$ are observed.
Relative pressure differences up to $4\times 10^{-3}$ were observed
at SE1 when operating the valve of the starting volume with full
closing speed.

Since the three expansion ratios at SE3 are determined from pressure
ratios which include the closing process, the effect of an enlarged
gas amount is reflected in the values of $f_{\mathrm{eff},i}$. It seems to
be advisable to reevaluate the value of $f_{\mathrm{eff},i}$ if the inlet valve
of the starting volume is replaced. The same valve closing speed
should be used during calibrations as was used to determine
$f_{\mathrm{eff},i}$. The minor contribution of the filling
pressure and gas-type dependencies are considered by a correction
factor and by a measurement uncertainty contribution, respectively.
Caution must be taken when transferring our observation to a different
valve. The values we present may serve as an estimate but cannot
replace an individual assessment of the valve. On the other hand, the
presented methodology is generally applicable.
These insights on the dynamic effects during the closing process of
small volumes pave the path towards an accurate assessment of enclosed
amounts of gas most relevant but not limited to the application in
static expansion systems.

\section*{Acknowledgment}
Kurt Sonderegger from VAT Group AG is acknowledged for the insights he
provided on the mechanical design of the valve.

\end{document}